\documentclass{sf2a-conf2025}
\usepackage{graphicx}
\usepackage{hyperref}
\usepackage[]{natbib}  
\usepackage{epstopdf}

\def\BibTeX{{\rm B\kern-.05em{\sc i\kern-.025em b}\kern-.08em
    T\kern-.1667em\lower.7ex\hbox{E}\kern-.125emX}}
\bibpunct{(}{)}{;}{a}{}{,}  

\begin{document}

\TitreGlobal{SF2A 2025}


\title{Enabling Blind and Visually Impaired Individuals to Pursue Careers in Science }

\runningtitle{Making Sciences more accessible}

\author{L. Petitdemange}\address{LIRA, Observatoire de Paris, Université PSL, Sorbonne Université, Université Paris Cité, CY Cergy Paris Université, CNRS,75014 Paris, France}

\author{S. Nashed}\address{Molecular Physicology and Adaptation, CNRS, Museum National d'Histoire Naturelle, 7 rue Cuvier, 75231 CEDEX 05 Paris, France}




\setcounter{page}{237}


\maketitle


\begin{abstract}

  Blind and Visually Impaired (BVI) Individuals face significant challenges in science due to the discipline's reliance on visual elements such as graphs, diagrams, and laboratory work. Traditional learning materials, such as Braille and large-print textbooks, are often scarce or delayed, while practical experiments are rarely adapted for accessibility. Additionally, mainstream educators lack the training to eﬀectively support BVI students, and Teachers for the Visually Impaired (TVIs) often lack scientific expertise. As a result, BVI individuals remain underrepresented in scientific jobs, reinforcing a cycle of exclusion. However, technological advancements and inclusive initiatives are opening new opportunities. Outreach programs aim to make science engaging and accessible for BVI individuals through multi-sensory learning experiences. Hands-on involvement in these activities fosters confidence and interest in scientific careers.
Beyond sparking interest, equipping BVI students with the right tools and skills is crucial for their academic success. Early exposure to assistive technologies enables BVI students to navigate scientific studies independently. Artificial Intelligence (AI) tools further enhance accessibility by converting visual data into descriptive text and providing interactive assistance. Several learning sessions demonstrated the eﬀectiveness of these interventions, with participants successfully integrating into university-level science programs. Educating BVI and their teachers on these tools and good pratices is the aim of our project AccesSciencesDV.

Research careers oﬀer promising opportunities for BVI, especially in computational fields. By leveraging coding, data analysis, and AI-driven tools, BVI researchers can conduct high-level scientific work without relying on direct visual observations. The presence of BVI scientists enriches research environments by fostering clearer communication, stronger collaborations, and more inclusive practices.

Despite these progress, systemic barriers persist. The lack of accessible educational resources, teacher training, and institutional support continues to hinder BVI individuals’ progress in science. Addressing these issues requires sustained eﬀorts in mentorship, technological development, and educational policy reform. In conclusion, by fostering inclusive learning environments, providing tailored support, and advocating for greater accessibility, contributing to the development of our project AccesSciencesDV, the scientific community can break down barriers and create a more diverse, innovative, and equitable field. Our experience shows that BVI individuals can become highly effective scientists when accessibility is systematically integrated throughout their educational and professional development.  
\end{abstract}



\section{Introduction}

Scientific disciplines present profound challenges for blind and visually impaired (BVI) individuals, largely due to their heavy reliance on visual representations. Fields such as physics, biology, and chemistry frequently use diagrams, graphs, and charts to convey complex concepts, creating significant barriers for BVI learners. Mathematics become particularly challenging due to the need for linearizing equations in Braille to make them comprehensible, often losing the structural clarity inherent in visual layouts. In addition, scientific learning often requires active participation in laboratory experiments and practical work, which are not accessible to BVI students. These barriers result in a persistent inequity that limits the engagement of BVI learners with science.
From an early age, BVI students face considerable difficulties in science education. It is important to note that the adaptations required vary greatly depending on the type and degree of visual impairment, whether the student is fully blind or has low vision. For partially sighted students, the effectiveness of adaptations often depends on the specific level of residual vision, making each solution highly individualized. The challenges faced by BVI students are partly attributable to the limited availability of adapted materials, such as large print and Braille-transcribed textbooks, as well as delays in their production, all of which significantly hinder the learning process \citep{BellS19}. This delay is compounded by the unavailability of tactile diagrams and raised-line drawings, which are essential for understanding geometric and spatial concepts \citep{RozenblumH15}. In mainstream educational settings, BVI students often struggle to follow board content, even when teachers attempt to verbalize mathematical problems. \citet{Archambault07} underscore the challenges of comprehending spoken mathematics, where ambiguities arise due to multiple possible interpretations of equations read aloud. Furthermore, access to specialized tools like Braille rulers for geometry and adapted calculators for algebra remains inconsistent. Practical work, often deemed too complex to adapt, is frequently omitted from the curriculum, limiting exposure to experiential learning that is critical for developing scientific skills \citep{OpieDS17, WestHTL04}. 
General education teachers are not at all trained to support students with visual impairments. Thus, support from Teachers for the Visually Impaired (TVIs) plays a crucial role, yet even this specialized assistance is often inadequate for addressing the specific needs of science education \citep{KoehlerW19}. TVIs often lack the expertise required to teach science at high levels as well as the knowledge on modern assistive technologies \citep{AjuwonGO16}. The same issues are observed in specialized instituttions such as the Institut National des Jeunes Aveugles (INJA) in France. that support BVI students in mainstream classrooms and offer a fully specialized in-house curriculum for those enrolled there.
  
The result of these barriers is a stark underrepresentation of BVI individuals in Science careers, creating a self-reinforcing cycle. The scarcity of BVI role models in science limits the aspirations of young BVI students, who may not perceive such careers as attainable. In France, for instance, their scientific opportunities are often limited to professions like physical therapy or software development, where specialized pathways exist. 
However, advancements in digital technologies and Artificial Intelligence (AI) open the door for BVI students to fully engage with scientific concepts and gain the skills necessary to pursue advanced scientific studies offering new career possibilities beyond traditional pathways.
As BVI scientists and educators, we present in this paper relevant strategies for addressing these challenges. Through outreach initiatives, we aim to inspire interest in science among BVI learners. By providing access to assistive tools and offering mentorship, we seek to equip them with the skills and confidence to pursue careers in STEM. Finally, we demonstrate how these efforts not only enable individual success but also contribute to a more inclusive and innovative scientific community.

\section{Igniting Scientific Curiosity: The Role of Outreach Initiatives }

\begin{figure}[ht!]
 \centering
 \includegraphics[width=0.8\textwidth,clip]{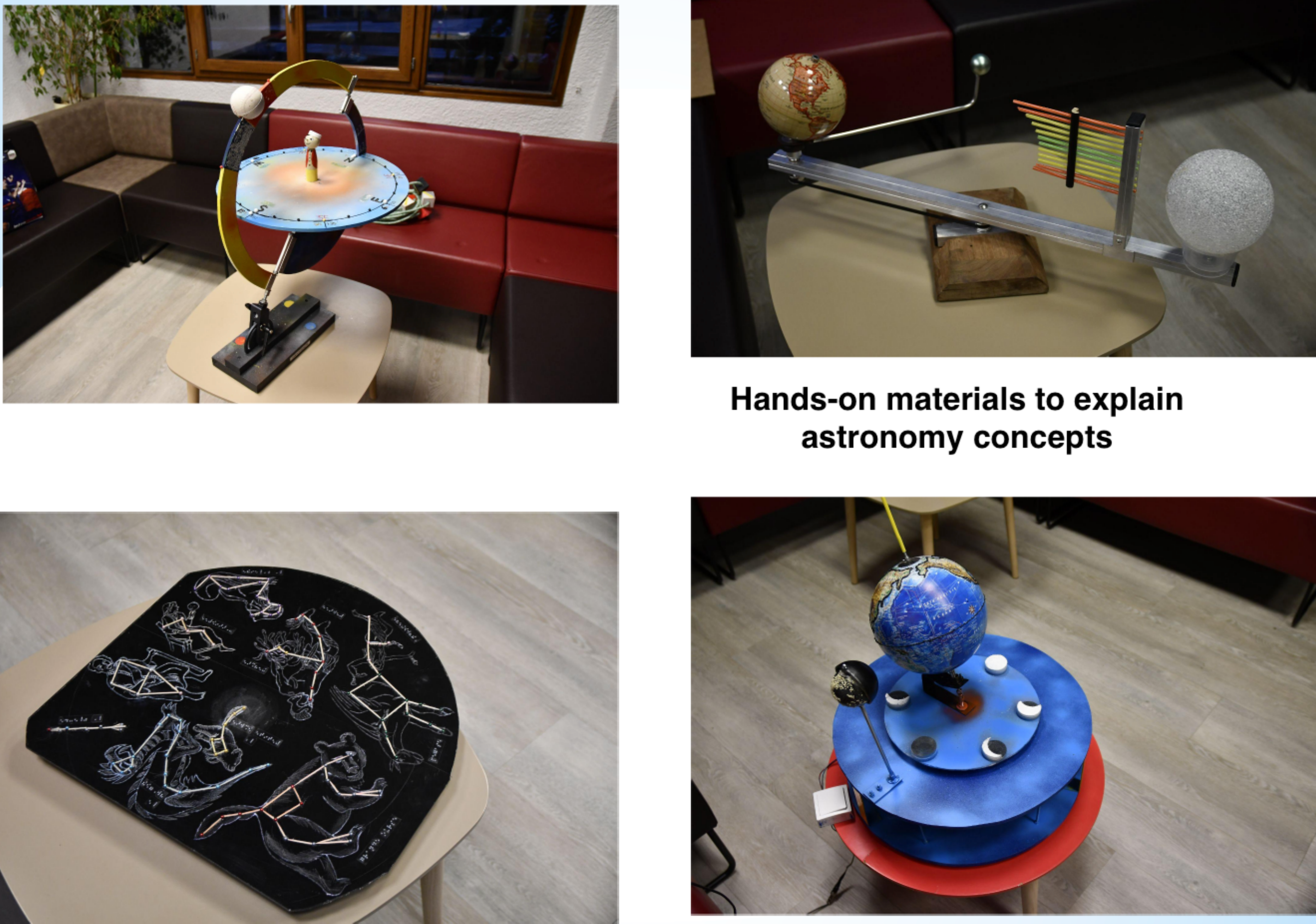}    
 \caption{Examples of adapted, hands-on educational materials enabling BVI and sighted individuals alike to grasp complex astronomical concepts such as the apparent motion of the Sun across the sky, the phases of the Moon, and the changing of the seasons. }
  \label{fig1}
\end{figure}

Outreach events play a pivotal role in introducing blind and visually impaired (BVI) individuals to the wonders of science. These initiatives go beyond merely making scientific content accessible; they aim to spark curiosity, inspire enthusiasm, and enable participants to envision themselves in scientific careers. Drawing on existing efforts to make astronomy accessible to BVI individuals \citep{BeckR08, Perez19}, we have developed unique activities in collaboration with the french Ciel d’Occitanie association (see Fig.~\ref{fig1}), which strives to make astronomy available to all.
These activities are designed to accommodate a diverse audience, including BVI individuals of all levels and degrees of visual impairment, as well as their sighted peers. By fostering an inclusive environment, these events promote mutual understanding: sighted participants gain insight into the lived experiences of BVI individuals, while BVI attendees benefit from an inclusive setting where social interactions are encouraged. Moreover, we posit that crafting explanations clear enough for BVI individuals inherently enhances their clarity for sighted participants as well. Grounded in a multi-sensory approach that incorporates sight, touch, and hearing, these activities offer to all immersive experiences that facilitate the comprehension of abstract scientific concepts. 
As BVI scientists, we are deeply involved in designing the scientific content used in these outreach events. This ensures not only that the material is fully accessible to the BVI audience but also that BVI individuals, including those running the activities, can actively engage in the dissemination of scientific knowledge. Acting as role models, BVI facilitators inspire younger generations by demonstrating that it is possible to achieve a high level of scientific expertise and share it effectively. Indeed, one of the key features of these events is their participatory nature. On specific occasions, young BVI participants are invited to co-host activities with us after attending our presentations. This hands-on experience allows them to develop scientific communication skills, build confidence, and feel empowered to engage in a field that is often perceived as inaccessible.
For blind individuals, touch is an indispensable sense for grasping complex scientific concepts. Our activities leverage this by employing specially designed tools and innovative technologies, such as 3D printing, to create tactile representations. We also design our own tactile models. For example, to represent the Milky Way, we use a model based on a magnetic plate shaped like the galaxy, covered with a thin sheet of cardboard. When sprinkled with iron filings, the particles adhere to the magnetic zones, providing a granular tactile representation that vividly conveys the density of stars in specific regions. This simple yet effective approach often elicits awe and fascination among participants. Another tactile model illustrates the relative distances between planets in the solar system. Large spherical beads, representing the planets, are strung along a taut cord. The spacing between the beads reflects the actual scaled distances between the planets, demonstrating, for instance, how much closer together the inner planets are compared to those farther from the Sun. We also use thermoformed tactile diagrams annotated in Braille, which have benefited from technological advancements that produce finer yet discernible tactile details. Additionally, we have developed a real-time observational device that captures celestial observations and produces tactile relief prints, offering an immediate and immersive experience. 
Hearing serves as another critical approach for understanding scientific concepts. All activities in our sessions are accompanied by precise verbal guidance and detailed descriptions, ensuring that participants can fully engage with the tactile models and other tools. During celestial observations, for example, participants receive verbal explanations of the relative sizes of stars, their distances from one another, and their colors, painting a vivid mental picture. Additionally, we employ auditory metaphors to convey abstract concepts; for instance, the spectrum of light is likened to a sound spectrum, with higher or lower frequencies corresponding to different wavelengths of light. One of our animations demonstrates this auditory approach by using a luminous sphere representing the Sun and an opaque sphere orbiting around it. Positioned in front of the luminous sphere is a camera that captures light intensity and converts it into sound. As the opaque object passes between the camera and the light source, the sound’s intensity decreases, offering a sonic representation of eclipses. This immersive setup enables participants to perceive the phenomenon of an eclipse not through vision, but through sound, highlighting the interplay between light and shadow in an accessible way.
While the proposed activities prioritize accessibility for BVI individuals, they also cater to sighted participants.
We organize astronomy discovery weeks that serve as an exceptional occasion for learning and exchange. These weeks are open to BVI participants of all generations and are structured around a diverse program. Several participants leave with newfound curiosity and enthusiasm for science. For us, these weeks also provide a valuable opportunity to gather feedback on our astronomy activities and improve them through a co-design approach, ensuring that our programs continuously evolve to better meet their needs. 

\section{Equipping BVI Students: Tools and Training for Academic Success}

Outreach events can spark an initial interest in science among BVI individuals, yet fostering this curiosity into a sustainable academic and professional pathway requires more than inspiration. For BVI students to engage meaningfully with science, they must be equipped with appropriate compensatory tools and skills early on, ideally during high school. As BVI scientists, we have firsthand experience of the challenges of navigating scientific studies without adequate preparation. We developed strategies on our own, and we now seek to share these insights with the next generation, serving as role models to inspire and guide them. This section explores the technological tools and skills that can empower BVI students to succeed in science and the importance of introducing these resources early in their academic journey.
Among the many tools that can be introduced early, the command line stands out as a particularly accessible and versatile resource for BVI individuals. It is entirely textual and linear, making it inherently compatible with screen readers such as the open-source NVDA or proprietary options, as well as Braille displays. Familiarizing BVI students with the command line allows them to realize that many software applications perceived as inaccessible due to their graphical user interface (GUI) can often be used effectively in command-line mode. This is particularly relevant in scientific disciplines, where the use of Linux is prevalent. However, many BVI students are primarily trained on Windows, which is more user-friendly for the general population. To bridge this gap, students can either learn to use Orca, the native screen reader for Linux, or take advantage of tools like Windows Subsystem for Linux (WSL), which enables Linux environments to run within Windows while maintaining compatibility with familiar screen readers and Braille displays. Although the command line may initially appear intimidating, introducing it early demystifies its use and opens up a wide range of possibilities. These skills enable BVI students to interact with scientific software, manage files, and execute complex analyses, forming a critical foundation for success in STEM fields.
The universal use of LaTeX for mathematical and scientific writing presents an invaluable opportunity for BVI students. Its text-based format naturally supports the linearization of equations, making it particularly suited to the needs of low-vision and Braille-using students. By introducing LaTeX in high school, BVI students can take notes directly in an accessible and structured format. Moreover, LaTeX is widely adopted by educators in scientific fields, facilitating direct communication between students and teachers and therefore eliminating the delays often caused by reliance on Braille transcription services. The integration of AI into tools like MatPix further enhances accessibility by converting PDFs into LaTeX, providing BVI students with readable, text-based versions of scientific texts. As Braille users are already accustomed to linearized content, LaTeX becomes a natural and efficient tool for interacting with complex mathematical structures. 
In addition to LaTeX, programming languages like Python or R in command-line (no graphic interface) provide another accessible and powerful alternative for tackling scientific challenges rather than using standard graphing calculators. The lack of accessible scientific calculators remains a significant barrier for BVI students starting in high school. Programming languages, however, offer a flexible solution. These languages are inherently text-based and linear, making them compatible with screen readers and Braille displays. In France, for example, Python is already introduced in high school mathematics classes, which positions it as an ideal starting point for BVI students. By deepening their understanding of Python, students can use it as a versatile tool—not only as a calculator but also for navigating datasets and creating visualizations. Programming languages thus empower BVI students to engage with scientific tasks on their own.
Beyond programming, recent advancements in artificial intelligence have expanded the possibilities for BVI individuals to access and interpret visual information. Tools like the paid version of ChatGPT and free applications like BeMyEyes now provide detailed, interactive descriptions of images, including scientific graphs and illustrations. These tools allow users to ask follow-up questions and gain deeper insights into visual data. As AI technology continues to evolve, its ability to generate precise and text-based descriptions will only improve, enabling BVI students to independently access and understand complex visual content.
While technology plays a crucial role in supporting BVI students, the importance of mentorship and role models cannot be overlooked. Beyond teaching technical skills, role models provide guidance on mastering assistive tools, collaborating effectively with teachers, engaging in group work, and maintaining personal motivation. By sharing their own experiences, role models demystify the path to success in STEM.
As a proof of concept, we organized a three-day summer school in August 2024 for BVI high school students considering higher education in science. The program had three participants and focused on teaching command-line usage, LaTeX, and Python, as well as advising on assistive technologies like using a Braille display in a terminal. Over the following three months, we monitored their progress. All three students successfully integrated into their university courses and actively utilized the tools introduced during the summer school. One participant now rarely relies on transcription centers, demonstrating the practicality of using LaTeX for independent scientific work. This initiative highlights that with proper training in compensatory tools—most of which are free and open-source—scientific studies, while challenging, can become a viable option for BVI students. 

\begin{figure}[ht!]
 \centering
 \includegraphics[width=0.8\textwidth,clip]{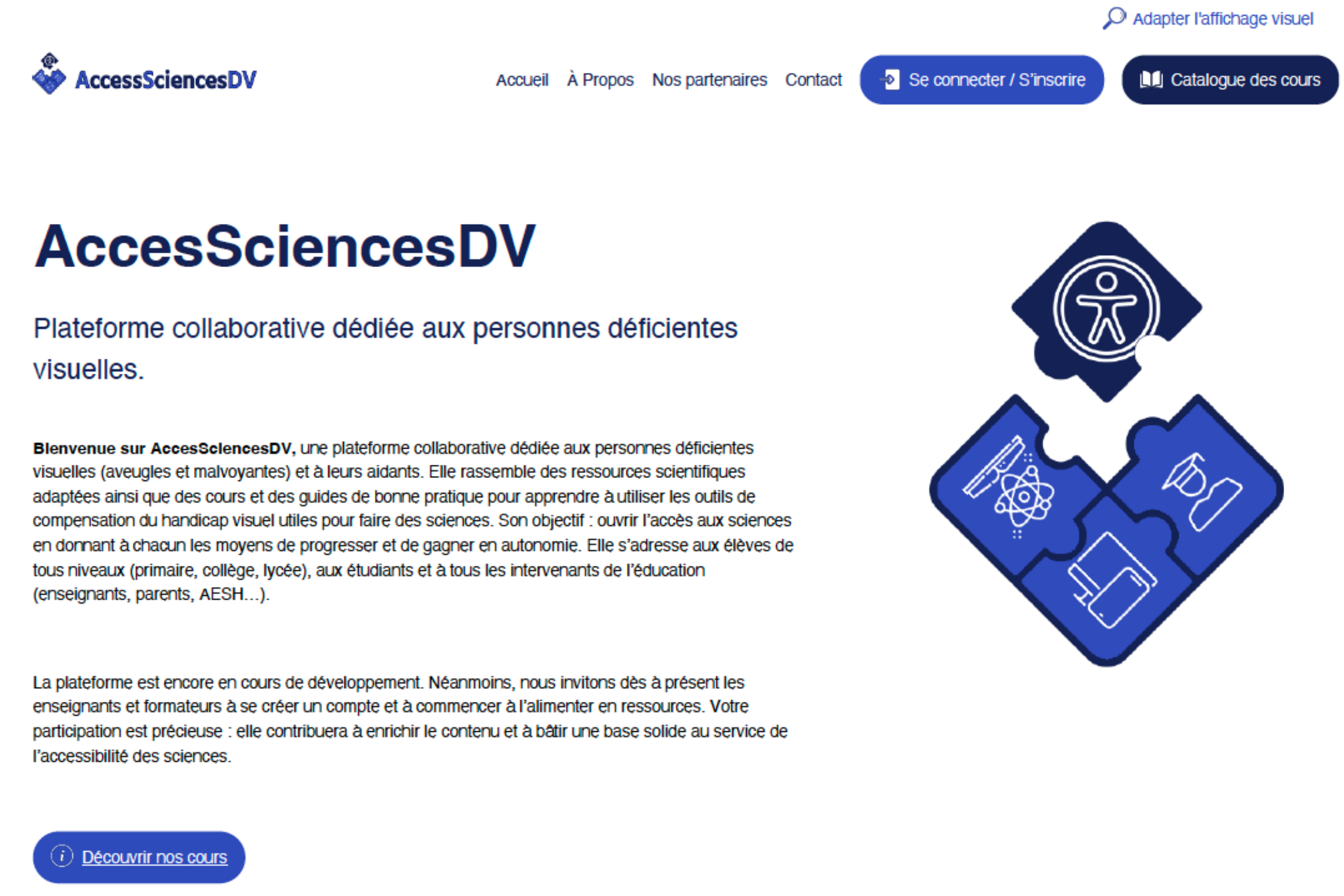}    
 \caption{ Picture of the website accessciencesdv.fr. The project AccesSciencesDV will collect accessible resources for BVI students. The collaborative elearning plateform give the opportunity to teachers and BVI students the good methods for making Sciences more accessible. }
  \label{fig2}
\end{figure}

In order to make such initiative more genral and accessible on the national level, we develop the project AccesSciencesDV accessciencesdv.fr. This accessible website will collect adapted resources in academic topics as: Mathematics, Biology, Physics\ldots and will provide the good pratices making Sciences more accessible for BVI students (see fig.~\ref{fig2})

\section{Integrating BVI Researchers: Challenges and Opportunities in Professional Science}

Blind and visually impaired (BVI) individuals who have successfully pursued higher education in science, equipped with the right tools and strategies, can fully integrate into professional scientific environments. Research careers, in particular, are becoming more accessible as many scientific fields increasingly rely on computational methods, which can be adapted to non-visual workflows. With the growing need for data analysis, modeling, and artificial intelligence-driven approaches, new opportunities are emerging for BVI researchers to contribute meaningfully across a wide range of disciplines. Our own experiences reflect this reality: one of us works as a postdoctoral researcher in bioinformatics, while the other has built a career as a research director in astrophysics. These achievements have been made possible largely thanks to accessible computational tools, which enable us to conduct data-driven research without direct visual interaction with laboratory experiments or telescopic observations.
Our successful integration into research relies on the same assistive tools discussed in the previous section. We utilize programming languages such as Fortran, Python, MATLAB, and R for data analysis, modeling, and visualization, allowing us to conduct our research autonomously. Moreover, LaTeX is an essential tool for writing scientific reports and publications, while Beamer, a LaTeX-based tool, enables us to create professional presentations for research conferences. Maintaining awareness of new assistive technologies is also a key part of our workflow. For instance, we have recently begun incorporating artificial intelligence to obtain feedback on the visual elements of our work, such as figures we generate ourselves or those found in research articles. This continuous adaptation ensures that we can work efficiently in a rapidly evolving scientific landscape.
Beyond simply adapting to our environment, we believe that our visual impairment can be a source of scientific innovation. The necessity of designing alternative ways to process and interpret data can lead to new methodologies that benefit the broader scientific community. A well-documented example of this is the case of a blind astrophysicist who made a discovery using sonification techniques, an approach that transformed complex astronomical data into sound, revealing patterns that were not apparent through conventional visual analysis (Díaz-Merced et al., 2019). More generally, the heightened level of rigor required in our work serves as an asset. Since we cannot rely on immediate visual feedback from our graphs, we must pay close attention to the underlying data, ensuring meticulous verification before drawing conclusions. This process minimizes the risk of misinterpretation caused by misleading visual patterns and fosters a deeper engagement with the data itself.
Furthermore, our presence within research teams contributes positively to their dynamics. Knowing that we cannot perform experiments in a laboratory or conduct direct astronomical observations, we frequently engage in co-supervision of students, creating strong collaborations. This involvement enhances teamwork and strengthens connections with colleagues. Additionally, our presence influences the way scientific discussions are conducted. In laboratory seminars, colleagues become more mindful of providing clear, descriptive explanations, which ultimately benefits all attendees by improving the clarity of presentations. Finally, our integration into research environments promotes inclusivity and fosters a culture of mutual support, demonstrating that scientific excellence and accessibility can go hand in hand.
Our personal experiences illustrate that with the right training, support, and adaptation strategies, BVI individuals can thrive in research careers. As scientific disciplines continue to evolve toward computational and data-driven methodologies, new opportunities will continue to emerge, ensuring that blindness or visual impairment does not preclude participation in cutting-edge research.

\section{Conclusion}

We have demonstrated that targeted interventions at different key moments can significantly improve the inclusion of BVI individuals in science. Outreach events play a crucial role in sparking interest and making science more engaging for BVI students. However, interest alone is not enough—providing the right tools and training early on, particularly in high school, is essential for enabling BVI students to pursue scientific studies on an equal footing with their peers. Furthermore, showing that scientific careers are achievable through computational approaches and accessible methodologies can help break the misconception that research is inherently inaccessible to BVI individuals.
Yet, while technology and inclusive education strategies are making scientific careers more feasible, this does not mean the path is easy. BVI students and professionals still face systemic barriers, from a lack of awareness among educators to the need for constant adaptation to evolving technologies. The scientific community must continue to push for greater accessibility, develop better assistive tools, and foster an environment of inclusion where BVI researchers can contribute fully.
The road ahead is still long, and challenges remain. However, we show that by intervening at multiple stages, equipping BVI students with the right tools, and promoting a culture of mentorship and adaptation, scientific careers are increasingly within reach. With the right support, BVI individuals can not only succeed in STEM but also enrich these fields with unique perspectives, creating a more diverse, innovative, and inclusive scientific community.

\bibliographystyle{aa}  
\bibliography{Petitdemange_S11} 

\end{document}